\documentclass[10pt,preprint,eqsecnum]{aastex}
\usepackage{amsmath}

\begin{document}

\title{Roche Lobe Overflow from Dwarf Stellar Systems.}

\author{Stephen D. Murray}
\affil{University of California,
Lawrence Livermore National Laboratory, P.O. Box 808, Livermore, CA 94550}

\author{Shawfeng Dong, \& Douglas N. C. Lin}
\affil{University of California,
Lick Observatory, Santa Cruz, CA  95064}


\begin{abstract}
We examine the evolution of residual gas within tidally-limited dwarf
galaxies and globular clusters.  In systems where the gas sound speed
exceeds about 10\% of the central velocity dispersion, hydrostatic equilibrium
causes the gas to have significant density at the tidal radius.  In contrast
to the stellar component, gas particles relax rapidly in response to mass 
loss.  Gas lost through the Lagrangian points is quickly
replenished, leading to continuous outflow, analogous to the
Jeans' escape of planetary atmospheres.  In such systems, the gas may be lost
on timescales as short as a few times the sound crossing time of the
system.  In colder systems, by contrast, the density contrast between the
core and the tidal radius is much larger, greatly reducing the mass loss
rate, and allowing the system to retain its gas for over a Hubble time.
Using this criterion as a guide, we present the results of numerical
simulations to show that either in the proximity of large galaxies or in
clusters of galaxies, residual gas in typical dwarf spheroidal galaxies
(dSphs) and globular clusters would be efficiently lost after it is ionized
and heated, either by external or internal sources.  The tidally removed gas
shall follow an orbit close to that of the original host system, forming an
extended stream of ionized, gaseous debris.  Such tidal mass loss severely
limits the ability of dSphs and globular clusters to continuously form stars.
The ordinary gas content in many dwarf galaxies is fully ionized 
during high red-shift epochs.  In extreme cases, therefore, star formation may
be completely prevented by early segregation of dark and ordinary 
matter, leading to the formation of starless, dark-matter concentrations
as the remnants of dwarf galaxy mergers in the CDM model of galaxy formation.
In the field, where the dwarf galaxies are tidally isolated, and in the center
of clusters of galaxies, where the tidal potential is compressive, 
ionized gas may be retained by dwarf galaxies, even though its
sound speed may be comparable to or even exceed the velocity dispersion.  In
such situations, dwarf galaxies may retain their gas content
until they merge with much larger galaxies.  These processes may help to
explain some observed differences among dwarf galaxy types, as well as 
observations of the haloes of massive galaxies.
\end{abstract}


\section{Introduction}
\label{sec:intro}

In the widely adopted cold dark matter (CDM) cosmological models,
small objects are the first to form, and larger galaxies form via the
merger of dwarf systems \citep[e.g.][]{WR78,BFPR84,C94,KNP97,NFW97}.
The evolution of the smaller building blocks is therefore of great
interest for the evolution of larger systems.  Because they
formed prior to the emergence of normal galaxies, these
subsystems are theoretically expected to have greater density than the
larger systems, and therefore they are likely to retain their
integrity in the halo of large host galaxies, which would imply the
existence of many more surviving systems than are observed today
\citep{Mooreetal99}.  The
lack of small stellar systems around larger host galaxies has been
referred to as the ``missing satellite'' problem \citep{Klypinetal99}.

One possible resolution of the problem is that gaseous baryonic matter
may have been ejected from the dwarf systems due to massive star
formation, which in turn
leads to photoionization-driven outflows, stellar winds, and supernovae
\citep{TBLN86,E92,LM94,QKE96,KBS97,NS97,WHK97,BL99,BLBCF02,DML03,
DS86,MLM88,MFR01, MFM02}.  Once
segregated, the ordinary matter continues to its dissipational infall,
while the dark matter components form a lumpy halo in the potential of
the large galaxy \citep{WEE98,SGV99}.  If the mass loss is driven by
bursts of star formation, a small amount of residual ordinary matter,
which has already been converted into stars, would remain with the clumpy
dark matter subsystems.  

In the halo of the Galaxy, several dwarf spheroidal galaxies (dSphs) are
found in the form of diffuse patches of stellar light \citep{Hodge71} embedded 
in dark matter haloes (Aaronson 1983; Faber \& Lin 1983).  It is natural to
identify these with the remnant subsystems, {\it i.e.} the theoretical building
blocks of large host galaxies in the CDM scenario
\citep{Mateo96,Klypinetal99,Mooreetal99}, 
because they have similar total mass ($\gtrsim 10^{7-8} M_\odot$).  There
are, however, significant differences.
Both the small (few km~s$^{-1}$) velocity dispersions, and the inferred
dynamical 
density of the observed dwarf galaxies are smaller than those postulated
in the CDM models.  Also, the spread in the observed amount of luminous matter 
(and hence the mass to light ratio) in the dSphs is much larger than 
that in their total mass \citep{Mateo98}.

Some of the above differences may be due to evolution.  The dSphs are found
to have a wide range of star-formation and enrichment histories, with few clear 
trends \citep{Mateo98,Grebel01,H01}.  All contain populations of old stars of 
varying sizes, and ages exceeding 10~Gyr.  In addition, some show histories 
of continuous star formation, many others show evidence of bursts of star 
formation separated widely in time, and a few show little or no evidence of
subsequent star formation.  The observed stellar population and the
modest mass-to-light ratio of the dSphs are consistent 
with the hypothesis that a significant fraction of the original baryonic
matter may have been ejected from them and other dwarf galaxies 
as a consequence of earlier epochs of star formation \citep{GCS83,vZSS98}.

In addition to dSphs, the families of low-mass, gas-poor dwarf stellar
systems also include dwarf elliptical galaxies, and globular clusters.  These
groups of objects are quite distinct from each other \citep{Kormendy85}.  Some
of their differences may, however, result from post-formation evolution
processes, such as mass loss either prior to or following star formation
(see discussion above), or segregation of baryonic and dark matter during star
formation \citep{MurrayLin89}.  The globular
clusters are the low-mass extreme of these families
of dwarf stellar systems.  Their velocity dispersions are generally more than
twice that of the dSphs within the Local Group of 
galaxies, but less than that of typical dwarf elliptical galaxies
within nearby clusters of galaxies.  In contrast to the dwarf spheroidal 
galaxies, all but one or two globular clusters of the Milky Way show evidence 
for only a single, old stellar population.  As in those dwarf galaxies 
which show little evidence for more than a single generation of star 
formation, any gas within globular clusters that remained after the 
initial generation of stars formed must have been lost from within them.  
Nevertheless, the multiple populations of stars in at least one globular 
cluster ($\omega$ Cen) and several dwarf galaxies indicate that the star 
formation process may not always be an effective mechanism for decoupling the
ordinary and dark matter in these loosely bound stellar systems.

Collectively, the population of the observed dwarf stellar systems falls far
below
that predicted in the CDM model, and so they alone cannot be the resolution
of the ``missing satellite'' problem.  Gravitational lensing
observations, however, find preliminary indications of starless dark
matter clumps in the halos of nearby galaxies \citep{Metcalf02, MZ02}.
If such dark entities dominate the halo structure, then an efficient and
severe segregation of ordinary and dark matter, through processes in
addition to the violent impact of star formation, is needed.  A
more quiescent process may therefore also play a significant role in
determining the ability of dwarf stellar systems to retain their gas to form
multiple generations of stars \citep{LM98}.  We shall address the cause
of multiple generations of stars in dwarf spheroidal galaxies in a
separate paper.

Several gas dynamical processes may lead to gas removal from stellar
systems with shallow potentials.  Prior to star formation, gas may be
removed by either ram-pressure or Kelvin-Helmholtz instability resulting 
from motion through the halo gas of a larger system \citep{LF83, MWBL93}.  
The efficiency of gas removal due to either process depends upon the
depth of the gravitational potential, and the density of ambient gas.  While
it is possible that the dwarf galaxies and globular clusters seen today were
able to retain sufficient gas to form stars within them because they have
deeper potentials than the dark matter-dominated ``missing satellites,'' 
the amount of ambient gas in the halo of the large, host galaxies is highly
uncertain.

In the halo of the Milky Way and other galaxies, both satellite dwarf 
galaxies and star clusters are subject to tidal forces.  
For dSphs, this process occurs following the 
formation of a nearby, massive galaxy, whereas star clusters form
within the potential of a massive system.  The tidal forces increase
rapidly in significance for smaller orbits around the host galaxy,
and can strongly perturb any ionized gas within the
satellite stellar systems.  The resulting situation is somewhat 
analogous to that of interacting binary stars \citep{Pringle85}, in 
that the ionized gas in the satellite systems always relaxes,
and expands to fill their Roche Lobes.  A fraction
of the gas continually streams away from the satellite systems through
their inner and outer Lagrange points.  If the rate of gas loss is
substantial, it may limit the ability of dwarf systems to retain gas
for subsequent generations of star formation.

In this work, we examine the significance of mass loss from
tidally-limited dwarf systems.  We begin in Section~\ref{sec:analytic}
with an analytic estimate of the rate of mass loss.  In
Section~\ref{sec:1Dmodels}, we examine the results of simple
one-dimensional models, and continue with three-dimensional models in
Section~\ref{sec:3Dmodels}.  Our results are summarized and
consequences for the evolution of dwarf systems are drawn in
Section~\ref{sec:discussion}.

\section{Analytic Determination of Mass Loss Rates}
\label{sec:analytic}

The distance of the L1 and L2 points from the center of the dwarf satellite
are given approximately by the Roche radius,
\begin{equation}
R_R=a\left({{M_d}\over{M_d+M_G}}\right)^{1\over3}
\left[{{1-e}\over{(3+e)^{1\over3}}}\right],
\label{eqn:xtapprox}
\end{equation}
for a point-mass parent-galaxy potential (King 1962), and 
\begin{equation}
R_R=a\left({{M_d}\over{M_d+M_G}}\right)^{1\over3}
\left\{ {{(1-e)^2}\over{[(1+e)^2/2e] {\rm ln} [(1+e)/(1-e)] + 1
}}\right\}^{1\over3},
\label{eqn:rroche}
\end{equation}
for a logarithmic parent-galaxy potential \citep{OhLin92}.  In the above
expression, $M_d$ is the total mass (gas, stars, and dark matter) of the
satellite system.  The parent galaxy has mass $M_G$ inside the orbital
semimajor axis $a$ of the satellite system.  
Although the concept of the tidal radius for a
satellite with an orbital eccentricity $e$ is uncertain and
controversial \citep{Innanen79, AllenRichstone88}, it provides a
useful fiducial length scale \citep{OhLin92}.  The above results may
be parameterized as
\begin{equation}
R_R=a\left({1\over\beta}{{M_d}\over{M_d + M_G}}
\right)^{1\over3}
\label{eqn:beta}
\end{equation}
where $\beta$ is a function of $e$. For circular orbits around a point mass
potential, $\beta=3$, whereas around an isothermal potential, $\beta=2$.

The rate of mass loss through the Lagrange points is related to the
Roche radius and the sound speed of the gas, $c_s$.  A full derivation
in the context of a low-mass planet orbiting around a main sequence
star is given in \cite{Guetal02}.  For the total loss rate of gas
through the L1 and L2 points, the results can be approximated as
\begin{equation}
\dot{M}_g=4\pi\rho_1 R_R^2c_s f,
\label{eqn:mdot}
\end{equation}
where $\rho_1$ is the gas density at the Roche radius.  The geometrical
correction factor, $f$, can be approximated by expanding the potential near 
the Lagrangian point (see \ref{eq:fest}).
 
If the gas within the dwarf system is isothermal, and the potential of
the system is approximated with a Plummer potential
\citep{BinneyTremaine}, then the gas density varies as
\begin{equation}
\rho_g(r)=\rho_0\exp\left[{{-\left(\phi(r)-\phi_0\right)}\over{c_s^2}}\right],
\label{eq:gasdis}
\end{equation}
while
\begin{equation}
\phi(r)={{-GM_d}\over{\left(R_c^2+r^2\right)^{1/2}}},
\label{eq:phiplu}
\end{equation}
where $\phi_0=\phi(r=0)=-GM_d/R_c$.  
In the above, $R_c$ is the core radius,
defining the radius over which the density of gravitating matter is
approximately constant. In this potential, the ratio of the gas density at 
$R_R$ to that at the center ($\rho_0$) is 
\begin{equation}
{\rho(R_R) \over \rho_0} = \exp\left[
{\alpha \over (1+x_R^2)^{1/2}} - \alpha \right]
\label{eq:f}
\end{equation} 
where $x_R=R_R/R_c$, and
\begin{equation}
\alpha\equiv {{GM_d}\over{R_c c_s^2}}.
\label{eqn:alpha}
\end{equation}
Even for relatively concentrated systems where $x_R \gg 1$, the gas
density retains a finite value at large radius, albeit
$\rho(R_R)/\rho_0$ can be very small for cold gas, with large values of
$\alpha$.  For large values of $\rho(R_R)$, self-gravity of the gaseous
envelope may become important, and the gas shall become a self-gravitating
isothermal sphere, resulting in a more rapid decline in density than
assumed here.

In response to mass loss, the gas adjusts on a sound crossing time in an
attempt to re-establish a density distribution appropriate to hydrostatic
equilibrium.  Any loss from the system at large radii is therefore
continually replenished.  This is in contrast to the stars, which diffuse on
the much-longer two-body relaxation time scale.  Thus, the mass loss rate
of the gas, $\dot M_g$ remains approximately constant prior to any
significant reduction in $M_g$.

The total mass of isothermal gas within the dwarf system is given by
\begin{equation}
M_g=4\pi\int_{0}^{R_R}\rho_g(r)r^2{\rm d}r = 4 \pi \rho_c R_c^3
\int_0 ^{x_R} {\rm exp}\left[{{-\left(\phi(r)-\phi_0\right)}
\over{c_s^2}}\right]{\rm d}x
\label{eq:mastot}
\end{equation}
where $x=r/R_c$. For a Plummer potential, 
\begin{equation}
M_g=
4\pi\rho_0 R_c^3\int_{0}^{x_R}\exp\left\{\alpha\left[{1\over{\left(1+x^2\right)
^{1/2}}}-1\right]\right\}x^2{\rm d}x.
\label{eqn:gasmass}
\end{equation}
Note that, for sufficiently large $x_R$, the dwarf galaxy's potential
can contain an arbitrarily large mass, even in the hot limit where
$\alpha<1$.  But for modest values of either $x_R$ or $R_R$, the
mass of gas within the potential is limited.

Using Equations~(\ref{eqn:rroche}), (\ref{eqn:mdot}), and (\ref{eqn:gasmass}),
and assuming the dwarf satellite to be on a circular orbit of period 
\begin{equation}
P=2\pi\left({{a^3}\over{GM_G}}\right)^{1\over2},
\label{eqn:period}
\end{equation}
the mass loss time scale becomes
\begin{equation}
\tau_{loss} \equiv{{M_g}\over{\dot{M_g}}} = 
{\Gamma\left(x_R,\alpha\right)P \over f},
\label{eqn:timescale}
\end{equation}
where
\begin{equation}
\Gamma\left(x_R,\alpha\right)\equiv
\exp\left\{\alpha\left[1-{1\over{\left(1+x_R^2\right)^{1\over2}}}\right]
\right\}
{1\over{x_R^3}}\left({\alpha\over{x_R}\beta}\right)^{1\over2}{1\over{2\pi}}
\int_{0}^{x_R}\exp\left\{\alpha\left[
{1\over{\left(1+x^2\right)^{1\over2}}}-1\right]\right\}x^2{\rm d}x
\label{eqn:gamma}
\end{equation}
gives the number of orbits of the dwarf satellite around the parent system
before significant mass loss occurs.

In the limit of cold gas, $\alpha\gg1$ and
\begin{equation}
\tau_{loss} \simeq
\left({{R_c c_s^2}\over{GM_d}}\right)
\exp\left\{ {{GM_d}\over{R_c c_s^2}}
\left[1-{1\over{\left(1+x_R^2\right)^{1/2}}}\right]\right\}
{P\over (2\pi \beta)^{1/2} x_R^{7/2} f}
\gg P
\end{equation}
so that over a Hubble time scale, no significant mass loss occurs.
But, for $\alpha<\alpha_1$, where 
\begin{equation}
{1 \over \alpha_1} \exp \left[\alpha_1 - 
{\alpha_1\over{\left(1+x_R^2\right)^{1/2}}}
\right] \simeq (2\pi \beta)^{1/2} x_R^{7/2} f,
\end{equation} 
a significant amount of gas is lost within an orbital period.  

\begin{figure}
\plotone{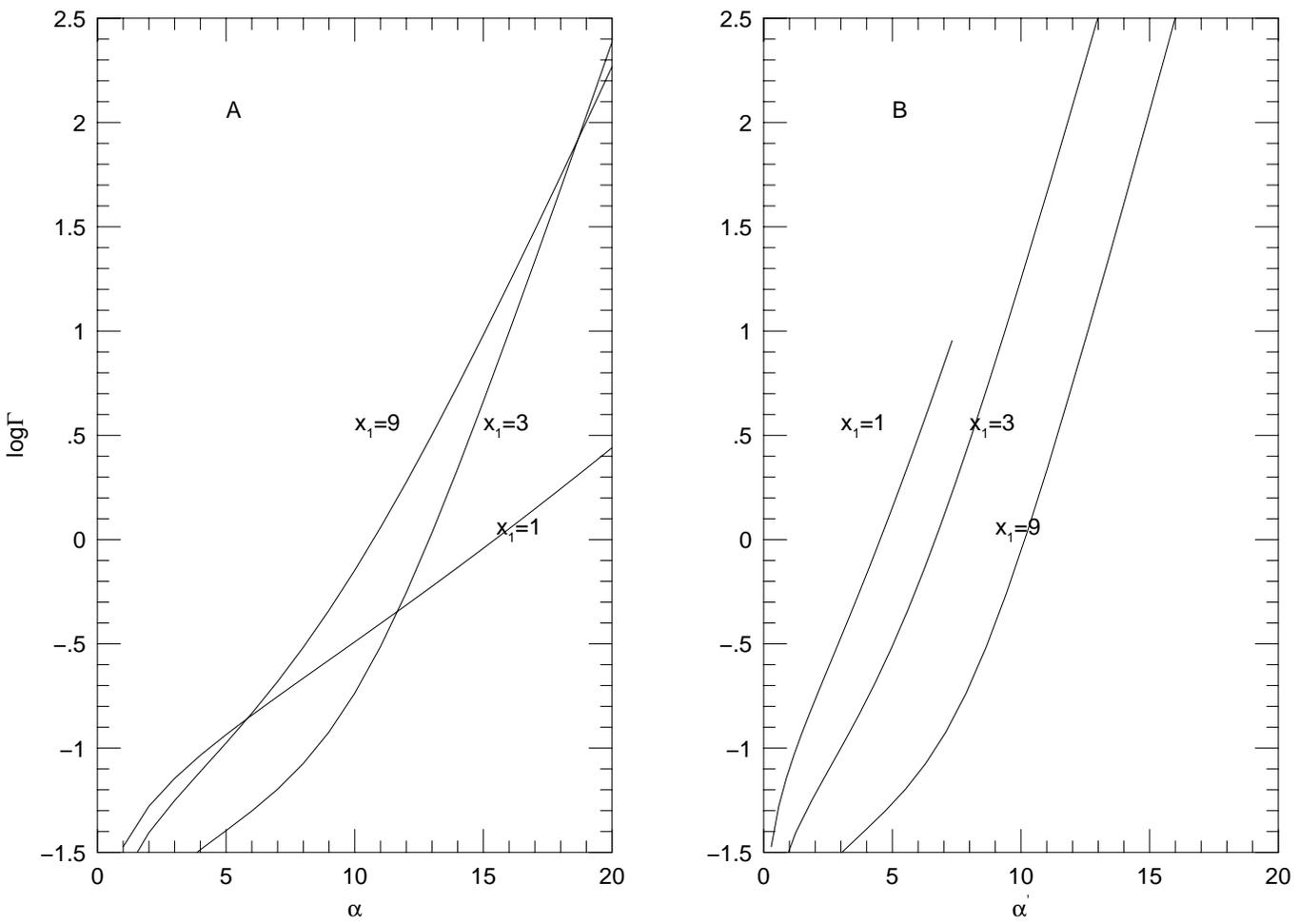}
\caption{The dependence of $\Gamma$ upon $\alpha$ (panel A) and upon
$\alpha^\prime$ (panel B) for $x_R=1, 3,$ and 9.}
\label{fig:gamma}
\end{figure}

In Figure~\ref{fig:gamma}, we show the variation of
$\Gamma\left(x_R,\alpha\right)$ with $\alpha$ (panel A) and
with $\alpha^\prime$ (panel B) for three values of
$x_R$, selected to match the values used in the one- and
three-dimensional dwarf galaxy models discussed in the next two
sections.  For these calculations, we choose $\beta=2$, corresponding
to a circular orbit in an isothermal host-galaxy potential.
The quantity $\alpha^\prime$ is defined as
\begin{equation}
\alpha^\prime \equiv \alpha \left[1 - \left(1 + x_R^2\right)
^{-{1\over2}}\right],
\label{eqn:alphap}
\end{equation}
such that ${\alpha^\prime}~^{1/2}$ corresponds to
the Mach number of the escape speed from the center to the Roche radius. 

As can be seen from panel A of Figure~\ref{fig:gamma}, and as would be
expected from Equations~\ref{eqn:timescale} and \ref{eqn:gamma}, the
time scale for loss of gas may become extremely short, less than the
orbital period, if $\alpha<\alpha_1 \sim10-15$.  The curves of
$\Gamma(x_R,\alpha)$ vs. $\alpha$ for $x_R = 1, 3,$ and $9$
cross each other for large $\alpha$.  We note, however, from panel~B of
Figure~1, that for $\alpha > 1$, $\Gamma(x_R, \alpha)$ is a more
sensitive function of $\alpha^\prime$
than of $\alpha$.
The critical condition for rapid mass loss, $\Gamma(x_R,
\alpha) =1$, occurs when $\alpha^\prime = \alpha_1^\prime \simeq 5-10$,
depending on the value of $x_R$.

The observed velocity dispersions of dwarf stellar systems, which are
indicative of the potential depths, are similar to
the sound speed of ionized gas
\citep{Mateo98,PM93}.  Once the gas in these systems is ionized,
either by internal or external sources, it shall have $\alpha^\prime \sim 1$.
In this limit, the results in Figure 1B imply that the ionized gas
may be rapidly lost by a continuous, quiescent outflow, thus
preventing the formation of subsequent generations of stars.
Subsequent stellar generations would form only if gas could be
accreted, a possibility which shall be examined in a subsequent paper.

\section{One-Dimensional Models}
\label{sec:1Dmodels}
In the analytic model presented above, the gas is assumed to be in
quasi-hydrostatic equilibrium at all times.  Yet, mass loss leads to
outflow and a reduction in the residual gas mass which should be
treated with a time-dependent dynamical calculation.  We examine the
time evolution of the gas in dwarf galaxies and star clusters using 
numerical simulations.  In order to compare with the analytic results 
above, we first carry out a series of one-dimensional dynamical 
simulations.  Using these models, we show the importance of ionization 
and gas temperature on the rate of tidally induced outflow.

The one-dimensional scheme used was first developed
to study star formation and feedback effects in dwarf
galaxies \citep{DML03}.  It is a 1D Lagrangian code, into which
radiative transfer and photoionization are incorporated using a
Str\"omgren shell model.  It thus allows us to explore the effects of 
heating and photoionization by both internal and external UV sources. 
With a Lagrangian scheme, we also do not need to specify artificial 
conditions near the boundaries of the computational domain.  Gas is allowed 
to freely flow out of the region.

For the 1D models, the gas experiences the gravity of the dark matter
of the dwarf galaxy, as well as an external tidal
potential.  The dark matter potential of the dwarf galaxy is taken to
follow the form found by Burkert (1995), which has been
tested to be a slightly better fit to dwarf spheroidal galaxies than the 
simpler Plummer potential, used in the previous section.  The
corresponding dark matter density distribution is given by
\begin{equation}
\rho_{DM} = \frac{\rho_c r_0^3}{(r + r_0) (r^2 + r_0^2)}.
\label{eq:rhodm}
\end{equation}
The central density and velocity dispersion of the
dwarf galaxies are scaled with a structural length scale
$r_0$ such that
\begin{equation}
\rho_c = 4.5 \times 10^{-2} (r_0/{\rm kpc})^{-2/3} M_\odot {\rm pc}^{-3},
\end{equation}
\begin{equation}
\sigma = 17.7 (r_0/{\rm kpc})^{2/3} {\rm km}~{\rm s}^{-1},
\end{equation}

The gravitational potential due to the dark matter within the dwarf galaxy
is determined from its density using Poisson's equation.
From equations (\ref{eq:gasdis}) and (\ref{eqn:gasmass}), we can determine
the density distribution of an isothermal gas in this potential, including 
its central value $\rho_0$.  In order to determine the extent of the dwarf
galaxy, we also impose an external tidal potential, chosen to follow the form
\begin{equation}
\phi_{tide} = \phi_0 \left[ 1 + \left( \frac{r}{r_0} \right)^{\lambda} \right]
\end{equation}
where $\phi_0 < 0$. The exact value of $\lambda$ is found be to unimportant; 
in the following calculations, we choose $\lambda = 1$. We specify the
tidal radius $R_R$ to be the location where the gradient of the total 
potential vanishes.  One can show that $R_R/r_0$ is related to $\rho_c$ 
and $\phi_0$ as
\begin{equation}
\phi_0 = - \frac{2 \pi G \rho_c r_0^4}{\lambda R_R^2} 
         \left( \frac{R_R}{r_0} \right )
         \left\{ \ln \left(1 + \frac{R_R}{r_0}\right) + \frac{1}{2} \ln
         \left[1 + \left(\frac{R_R}{r_0}\right)^2 \right] -\arctan 
         \left(\frac{R_R}{r_0} \right) \right\}
\label{eq:phi0}
\end{equation}

The initial distribution of gas is set up in hydrodynamic equilibrium
under the dark matter potential alone. The total mass of gas,
$M_{g} = q M_{DM}$, where the ratio of baryonic to dark matter $q$ is
assumed to be 0.1 in most of our calculations \citep{Mateo98}.  
At the onset of the calculations, both the external potential and 
extragalactic UV radiation are switched on instantly . The residual
gas quickly settles into a quasi-equilibrium after a brief and minor 
adjustment in the outermost region.  Because the gas density is
tenuous in this region, this adjustment does not affect the subsequent 
evolution of the gas.  

We assume the gas to be metal-free, so that in the absence of
photoionization it would cool to a temperature $T\sim 10^2K$
\citep{KSFR90, LM92}.  The temperature of the photoionized 
gas is taken to be 15,000~K, appropriate for low-metallicity,
photoionized gas.  The external UV flux
$J_{\nu}\approx10^{-23}$~ergs~s$^{-1}$~cm$^{-2}$~ster$^{-1}$~Hz$^{-1}$ at the
Lyman limit at low redshift, and $50$ times more intense at redshift $z=3$
(cf. Haardt \& Madau 1996), is sufficient to completely ionize
most dSphs, as confirmed by our models.  More massive dwarf elliptical
galaxies (dEs), are, however, partially self-shielded from the external
UV, and so they contain a region of neutral residual gas surrounded by ionized
gas.  Because we are, in this section, primarily interested in the efficiency
of photoionization for dEs, the model parameters used here are appropriate
to those systems.  In the next section, we present the results of some
3-D models for the evolution of fully ionized gas within dSphs.

The properties of the one-dimensional models are shown in
Table~\ref{tab:1Dgalaxies}.  For each model, we list $M_d$, 
$r_0$, the escape speed from the central potential ($\sigma$,
equivalent to $GM_d/R_c$ in the previous section), $\alpha^\prime$ for the
ionized gas, $J_\nu$, the initial baryonic mass fraction
($M_{gi}/M_d$), the ratio of the final mass of neutral gas to the initial total
gas mass ($M_{Hf}/M_{gi}$), the ratio of the initial radius of the sphere of
neutral gas to the Roche radius ($R_R/R_n$), $P$, the computed
mass loss timescale for gas ($\tau_{loss}$), the ratio of $\tau_{loss}$ to $P$,
and the computed value of $\Gamma$.

In the standard model (1D-1), $r_0 = 1$ kpc,
such that $\sigma \simeq 18$~km~s$^{-1}$.  In this and
all other models presented here, we set $R_R=3 r_0$, so that the dwarf
galaxy's total mass $M_d \simeq 5 M_0$ where $M_0$ is the total mass within
$r_0$.  In Model 1D-1, the galaxy's mass (including both dark and
gas components) within $r_0$ and $R_R$ are, respectively,
$M_0 = 7.2 \times 10^7 M_\odot$ and $M_d=3.66 \times 10^8 M_\odot$.

If the dwarf galaxy has a circular orbit in an isothermal potential of
the background host galaxy, with circular speed
$V_c=210$~km~s$^{-1}$, and mass
\begin{equation}
M_G = 5 \times 10^{11} (a/50 {\rm kpc}) M_\odot,
\end{equation}
the orbital radius $a$ is given by Equation~(\ref{eqn:beta}).  The orbital
period, $P=1.49\ (a/50\ {\rm kpc})$~Gy and $\beta=2$. For Models
1D-1, 1D-2, 1D-3, and 1D-4, we adopt $r_0 =1$~kpc so that 
$a=129.1$~kpc and $P=3.85$~Gy.
For Model 1D-5, we choose $r_0=1.35$~kpc, such that $a=182.6$~kpc
and $P=5.14$~Gyr.

\begin{figure}
\plotone{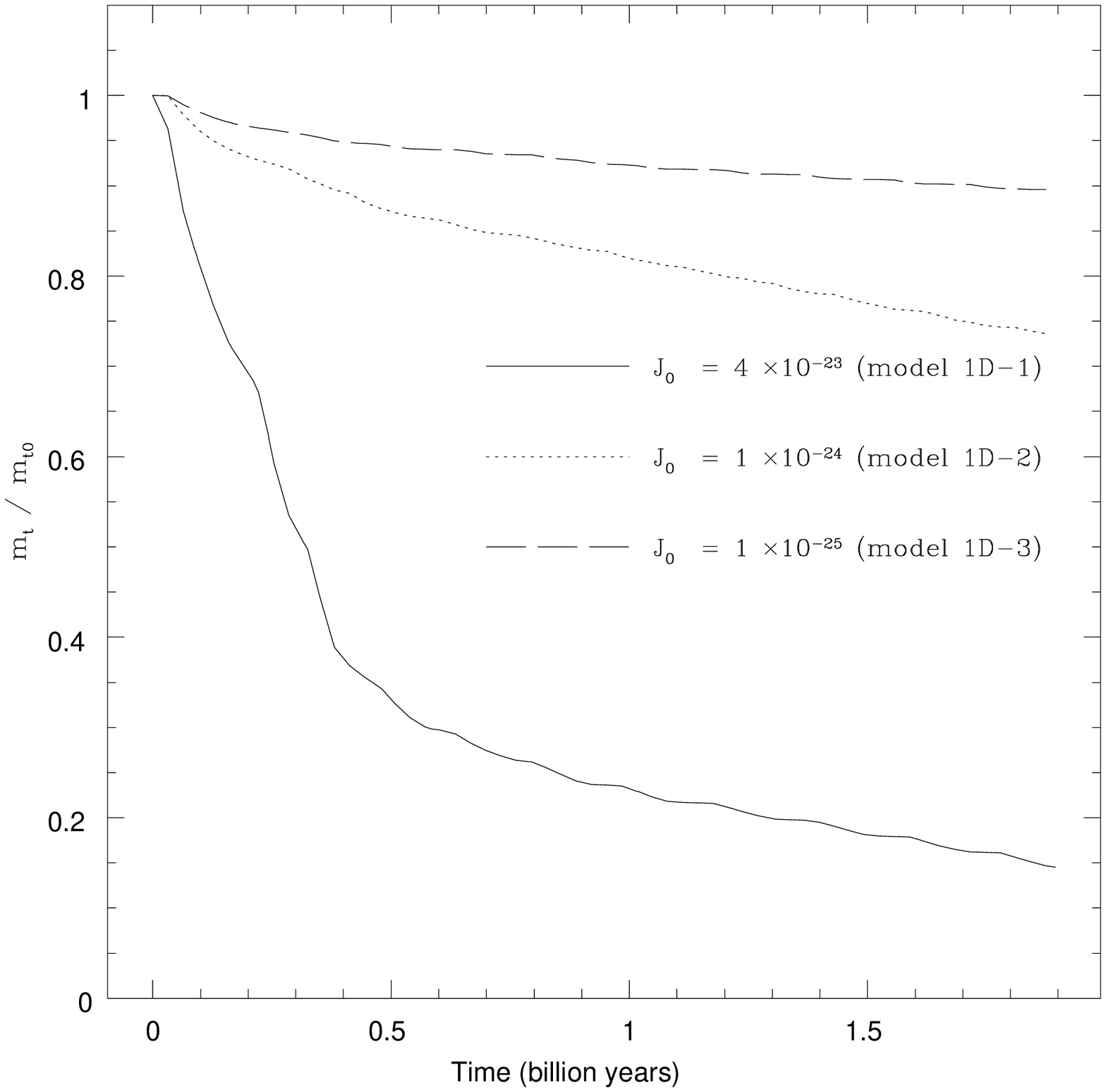}
\caption{Mass of gas contained within the tidal radius as a function
of time for the one-dimensional models 1D-1, 1D-2, and 1D-3.}
\label{fig:1Dtidea}
\end{figure}

\begin{figure}
\plotone{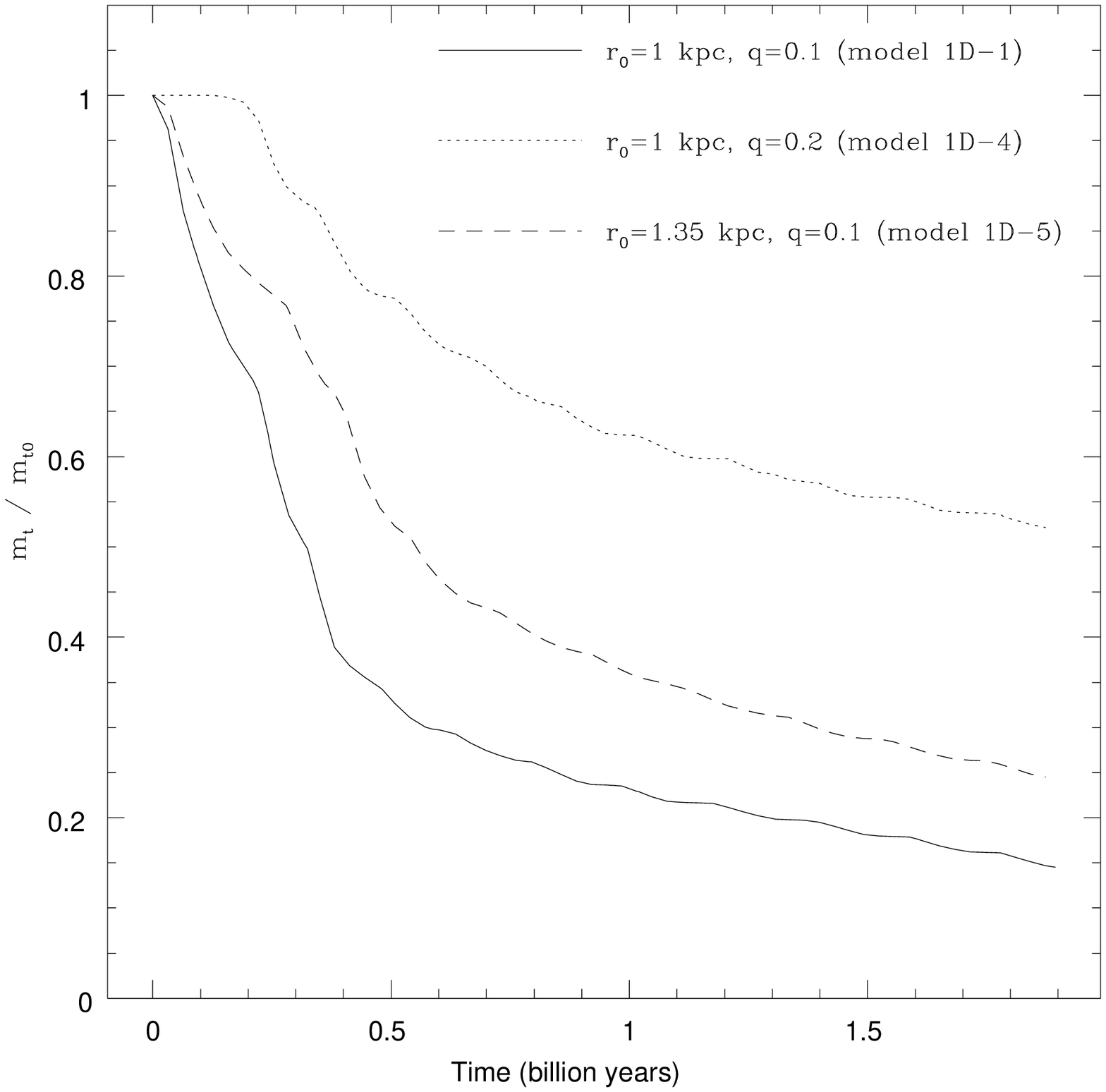}
\caption{Mass of gas contained within the tidal radius as a function of time
for the one-dimensional models 1D-1, 1D-4, and 1D-5.}
\label{fig:1Dtideb}
\end{figure}

The evolution of the mass fractions of gas within the tidal
radius for the one-dimensional models are shown in Figures~\ref{fig:1Dtidea}
and \ref{fig:1Dtideb}.
As can be seen in Figure~\ref{fig:1Dtidea}, half of the gas mass is lost
from Model~1D-1
within 350~Myr, in spite of the fact that $c_s \simeq 0.6 \sigma$.  This
result is consistent with our discussion in the previous section where
we showed that a significant fraction of the ionized gas is lost
within one galactic orbital period for $\alpha = (\sigma/c_s)^2 <
\alpha_1 \simeq 10$.  Such rapid outflow of gas would prevent extended
and protracted star formation.
The ionized gas at the outskirts of the galaxy flows beyond $R_R$ with
the sound speed ($\sim 10$~km~s$^{-1}$) of the ionized gas.

\begin{figure}
\plotone{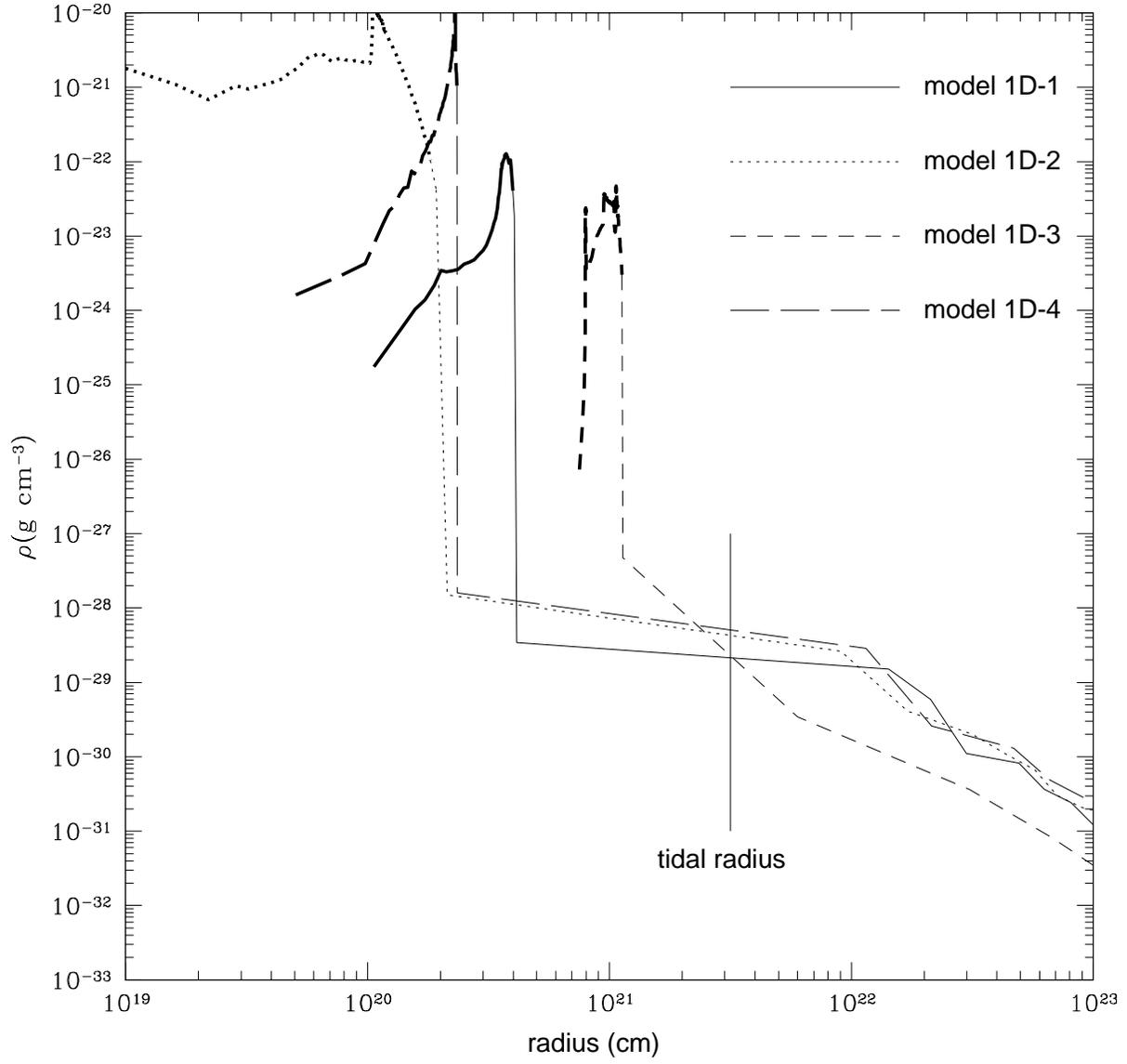}
\caption{Density of gas for the one-dimensional model. The thicker lines
denote the density of neutral gas $\rho_n$, while the thinner ones denote the
density of ionized gas $\rho_i$.}
\label{fig:1Drho}
\end{figure}

In Model~1D-1,
$J_{\nu} = 4 \times 10^{-23}$~ergs~s$^{-1} $~ cm$^{-2}$~ster$^{-1}$~Hz$^{-1}$,
appropriate for redshift $z=1$.  In this model, a small
amount, $M_{Hf}$, of the residual gas remains in a neutral state at the
center of the galaxy, surrounded by ionized gas at an interface of radius
$R_n$.  In Figure~\ref{fig:1Drho} are shown snapshots of the densities of
the neutral and ionized gas, $\rho_n$ and $\rho_i$ for the first four
one-dimensional models, as a function of radius.  These results
show that, analogously to the Str\"omgren sphere in an HII region, the
ionization front contracts slowly from the outer tenuous to the inner, denser
regions of the galaxy, while the ionized gas lost from the outer halo
is being continually replenished. 

When a quasi-static equilibrium is established, in which the external
UV flux is balanced by recombination in the outer ionized
envelope, the magnitude of $\rho_i \propto J_\nu ^{1/2}$
(Figure~\ref{fig:1Drho}), while $c_s$ is independent of $J_\nu$.  We thus
expect $\dot M \propto J_\nu^{1/2}$.  This conjecture is confirmed by 
Models 1D-2 and 1D-3 which have identical structural parameters as 
Model 1D-1, but reduced $J_{\nu}$.  In these cases, both the mass
fraction and radius of the neutral sphere are larger than the standard model.
Similar to the standard model, $\alpha \simeq 1 < \alpha_1$ in the ionized
envelope and the ionized gas flows from the ionization front at $R_n$ to $R_R$
in a fraction of the sound crossing time.  The density of ionized gas at
$R_R$ is, however, reduced relative to the standard model, due to the larger
masses of cold, neutral gas, with its steeper radial density falloff.  The
result is a reduction in $\dot M_d$ relative to the standard model.
The actual dependence of $J_{\nu}$ upon redshift remains uncertain,
and the values listed in Table~\ref{tab:1Dgalaxies} are comparable to the
values estimated for the present epoch by some authors
\citep{HM96,BWLM87,OI83}.

In Model 1D-4, we increase the ratio, $q$, of the baryonic-to-dark
matter mass by a factor of two from the standard model.  The
increased density of baryonic matter enhances self-shielding,
and the gas becomes mostly neutral.  The gas cools and settles to
the center of the galaxy, with a smaller $R_n$.  The ionization rate
across the front is balanced by rapid recombination, making
photo-evaporation less efficient.  Consequently, the value of $\rho_i$
decreases, and the mass loss rate is reduced (Figure~\ref{fig:1Dtideb}).
In Model~1D-5, we increase $r_0$ by a factor of 1.35 but maintain the
baryonic fraction at 0.1.  In accordance with the Burkert (1995)
potential, $M_d$ in Model~1D-5 is twice that in Model~1D-1, though the
initial gas mass $M_{gi}$ is the same as Model~1D-4.  
With these parameters,
gas in the center of the dwarf galaxy is less dense, but the total
baryonic mass is greater than in the standard model.  Once again,
the larger gas reservoir enables a greater fraction 
of $M_{gi}$ to become
self-shielded, neutral, cool, and settle to the center of the
galaxy, again reducing the mass loss rate relative to the standard model.

We have also computed a case which has identical physical parameters as
Model~1D-1, but the galaxy is set in isolation (infinite tidal radius).  In
this case, the gas adjusts to both ionization and
hydrostatic equilibrium with no gas attaining finite outflow velocity.  As
indicated above, even for the early epochs where $z$ and $J_\nu$ are large
and the gas is fully ionized and hot, these static equilibria can be attained
from Equations~(\ref{eq:gasdis}) and (\ref{eqn:gasmass}) with a 
sufficiently large $x_R$ or equivalently $R_R$. The 
implications of these results will be discussed in \S~\ref{sec:discussion}.

\section{Three-Dimensional Models}
\label{sec:3Dmodels}

The one-dimensional models provide the basic description for the
underlying ionization, relaxation, and hydrodynamics of the residual 
gas which regulate the efficiency of its tidally induced
outflow.  But in reality, the tidal forcing of the host galaxy is
non-spherically symmetric about the center of the dwarf galaxy.  In order
to consider a more realistic tidal potential, we examine the results of
some three-dimensional simulations in this section.

\subsection{Numerical Method}

The numerical code we use is Cosmos.  It is a massively
parallel, multidimensional, radiation-chemo-hydrodynamics code
developed at Lawrence Livermore National Laboratory.  The code, and
tests of the code are described in \cite{AF03}, and \cite{AFM03}.  It
has also been used to study the evolution of supernova-enriched
material in dwarf galaxies by \cite{FMAL03}.  The reader is referred
to those papers for more details on Cosmos.  We discuss here the
settings used for the current work.

We wish to examine the evolution of either cool or heated gas within
dwarf galaxies and massive star clusters.  The one-dimensional
models discussed above include the effects of heating, cooling, and
ionization of
the gas and show that the gas dynamics is primarily determined by its
temperature and more indirectly by the state of ionization. For the
three-dimensional models, the inclusion of ionization and recombination
of the gas would only reduce the controlled nature of the models,
without adding new information.  Thus, we choose to set the gas to be
initially isothermal, at a determined temperature.  In order to
approximate the effects of heating and cooling, which has the tendency
to maintain the gas at a nearly constant temperature, we assume that
the gas evolves adiabatically, with exponent $\gamma=1.1$.  We choose
two sets of models.  For the hot models, we set the gas temperature to
be 8,000~K and for the cold model we assume the gas is in an atomic
state with a temperature of 1,000~K.  These isothermal states
correspond to either early or recent epochs, when $J_\nu$ is either adequate
or insufficient to significantly ionize the residual gas.  The results of
the 1-D models in the previous section indicate that, for dSphs, the entire
galaxy is likely to be ionized by a modest external UV flux. 
Thus an isothermal temperature is a reasonable approximation for the hot 
models.  When the gas attains a sufficiently large density, it becomes
self-shielded inside these galaxies and clusters.  The isothermal approximation
is also sufficient for this cool phase.

Self-gravity of the gas is not included in the models.  This
approximation should be adequate for dwarf galaxies, which are
dark-matter rich, and so their gravity is dominated by that of the
dark matter.  For a globular cluster, which has undergone efficient
star formation, the potential is dominated by the stars, and so,
again, the potential of the gas is not important to the evolution.
The models therefore include a static background potential, within
which the gas evolves.  In order to compare the numerical results with
our analytic calculations, the background is modeled as a Plummer
potential, as expressed in Equation~(\ref{eq:phiplu}), which has a form
similar to, but simpler than, those found to be followed by dwarf
galaxies \citep{B95}, and which has also been used to
model globular clusters \citep{Plummer11}.

To the potential of the dwarf stellar system is added that of a
massive parent galaxy.  The total density of the massive parent galaxy
is assumed to follow the same isothermal form as above, with a total
mass of $5\times10^{11}$~M$_\odot$ and a circular speed $V_c = 210$ km
s$^{-1}$ at some typical distance 50~kpc from the Galactic center.
Smaller Galactic radii are used for the globular cluster models.

Throughout the numerical simulations, our models remain centered upon
the dwarf systems, ie. we are in a non-inertial frame of reference.
Coriolis terms must therefore be added to the potential, which then
has the form
\begin{equation}
\phi(x,y,z)=\phi_d(x,y,z) + \phi_G(x,y,z) -
{1\over2}\Omega^2\left(x^2 + y^2\right),
\label{eqn:rochepotential}
\end{equation}
where $\phi_{d,G}$ are the conservative gravitational potentials of,
respectively, the dwarf stellar system and parent galaxy, $\Omega$ is
the angular velocity of the rotating frame (the system is assumed to be
in a circular orbit), the $z$-axis is the axis of rotation, and the
$x$-axis points towards the center of the parent galaxy.  This form of
the effective potential is analogous to that used for close binary
stars \citep{Pringle85}.

\begin{figure}
\plotone{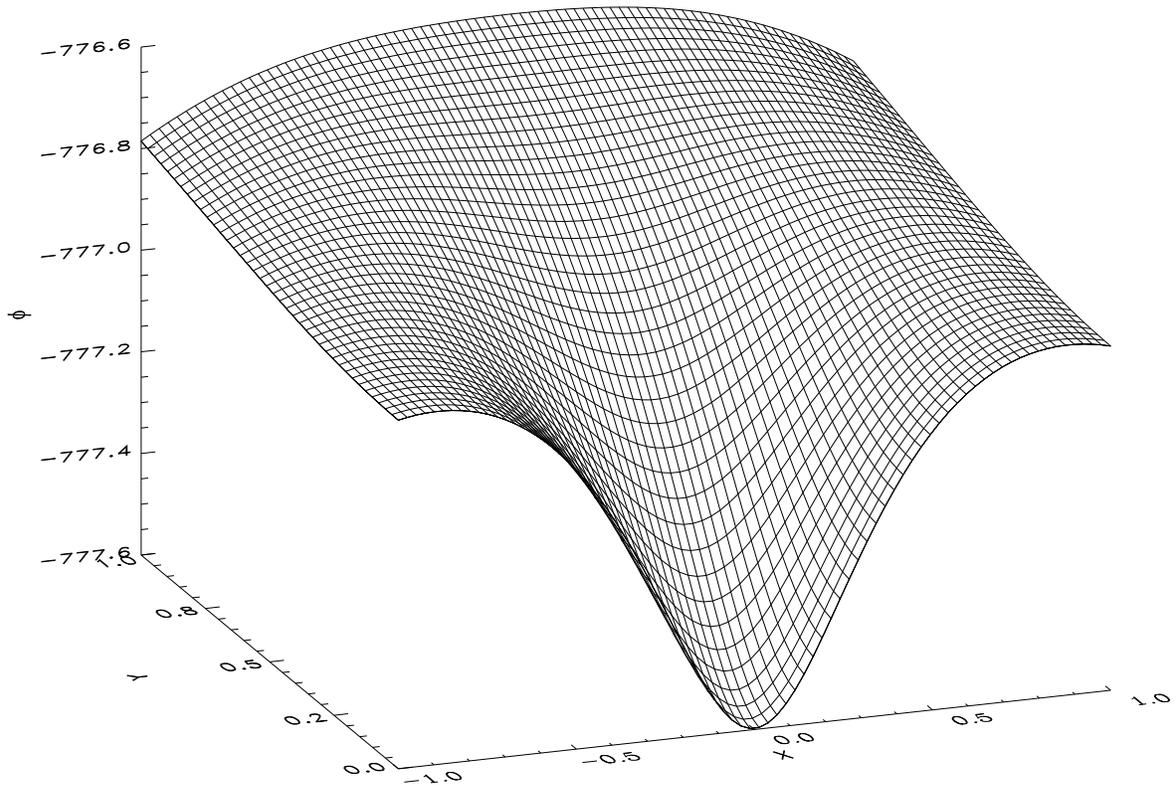}
\caption{Surface plot of $\phi(x,y,z=0)$ for the three-dimensional
galaxy Model~3D-1d.  In code units, the $y$-axis runs from 0 to 1, the
$x$-axis from -1 to 1, and $x_t=0.89$.  The potential, $\phi$, is
in code units.
}
\label{fig:rochepotential}
\end{figure}

The form of the potential is shown in Figure~\ref{fig:rochepotential},
which shows a surface plot of $\phi(x,y,0)$ for Model~3D-1d.  Positions
on the $x$ and $y$ axes, and the potential are all in code units.
The L1 and L2 points are located along the $x$-axis, at $x=\pm x_t$, 
where $x_t=0.89$ in code units.  Due to the saddle-shape of the potential
at L1 and L2, mass loss may occur only near the
Lagrange points.  At an orbital radius $a$ where $M_d \ll V_c^2 a/G$
\begin{equation}
x_t = \pm \left[ \left({ G M_d \over 2 \Omega^2 R_c^3} \right)^{2/3}
-1 \right]^{1/2} R_c
\end{equation}
which reduces to $x_t \simeq (G M_d/ 2 \Omega^2)^{1/3}$ in the
limit that $R_c \ll (G M_d / 2 \Omega^2)^{1/3}$.  In comparison with
our analytic, spherically symmetric approximation, $x_t$ is the analogous
quantity to $R_R$, and is only
slightly modified from the expression for the Roche radius of a
point-mass potential for the dwarf galaxy.  

In the vicinity of $x_t$,
the gas density is approximately constant because the gradient of
$\phi$ vanishes, such that the density scale height of the gas diverges
at $x_t$.  The escaping gas streams out of the dwarf galaxy through the
saddle surfaces, which are bounded by the local potential maxima in
which $\partial \phi/\partial x =0$.  This surface lies approximately
on the $y-z$ plane, centered on $x=x_t$.  For $x_t > G M_d/C_s^2$, 
we can estimate the magnitude of $f$ (cf. Equation~\ref{eqn:mdot}) under 
the assumption that gas diffuses to the outflow channels in a hydrostatic 
manner, such that
\begin{eqnarray} 
f & \sim & {1 \over 4 \pi x_t^2} \int_{-\infty} ^\infty \int_{-\infty} 
^\infty {\rm exp}\left[- {\phi(x_t, y, z) - \phi(x_t, 0, 0) \over
c_s^2 }\right]
dy dz \cr \nonumber \\
& \sim & {x_t c_s^2 \over \pi G M_d} \int_0 ^\infty {\rm exp}\left(-{3 z^2
\over 4}\right) dz \int_0 ^\infty \left(-{y^2 \over 2}\right) dz =
\left( { 2 \over 3} \right)^{1/2}
\left({x_t c_s^2 \over G M_d}\right) \sim {x_R \over \alpha}.  
\label{eq:fest}
\end{eqnarray}
Although hydrostatic equilibrium is approximately maintained within 
the tidal radius, the outflow via the channels prevents the gas from 
establishing a sustained equilibrium, and that the magnitude of $f$ may 
be quite different from the above estimate.  Nevertheless, the above 
approximation suggests that $f$ is unlikely to be orders of magnitudes 
smaller than unity for modest to large values of $\alpha^\prime$.
For $\alpha <1$, $f$ in Equation~(\ref{eqn:mdot}) cannot exceed unity.

In principle, non-inertial terms should also be added to the equations
of motion of the gas.  These would, however, strongly affect only that
gas which was either at or beyond the tidal radius.  Because our goal
in this work is to examine the rate of mass loss, and not the
subsequent evolution of gas lost to the dwarf system, omission of the
non-inertial terms shall not affect our conclusions significantly.
Our models are generally integrated for a small fraction of an orbit,
and so their affect would be small in any event.

The gas density is initially set up such that
\begin{equation}
\rho(x,y,z)=\rho_0\exp\left[-\phi\left(x,y,z\right)/c_s^2\right],
\label{eqn:initialdensity}
\end{equation}
where $c_s^2$ is the isothermal sound speed of the gas.  For a
spherically-symmetric potential, such an initial distribution of
isothermal gas would describe hydrostatic equilibrium.  For the
non-spherical potential described above, however, the density of the
gas near $R=x_t$ evolves significantly at early times.  In order to
prevent this from causing us to overestimate the mass loss rates, we
calculate the mass of gas contained within $R<x_t/2$, which we denote
as $M_2(t)$.  That quantity shows little affect from the behavior at
larger radii.  In addition, we avoid influences from early behavior by
fitting the value of $M_2(t)$ at late times.

Cosmos is written for Cartesian coordinates, and so we cannot take 
advantage of the symmetry of the potential to run the problem in
two dimensions.  In order to reduce computation time, the problems are
run as quadrants.  Typical resolutions are 200x100x100.  We have
examined the effects of changing resolutions in some of the models, and
find that doubling the resolution changed the time scales for mass loss
by only 10\%.

\subsection{Model Results}
\label{sec:3Dresults}

The models which we have run in order to examine the evolution of gas
within dwarf galaxies are shown in Table~\ref{tab:3Dgalaxies}.  On the
table are listed, for each model, the total mass of dark matter in the
dwarf, $M_{dm}$, the core radius of the dark matter potential, $R_c$,
the depth of the potential, $\phi_0^{\prime 1/2}$ (measured from the
center to the galaxy to $x_t$ along the y-axis, i.e. $\phi_0 \left[1-
(1+x_R^2)^{-1/2}\right]$, the orbital radius of the dwarf galaxy around the
parent system, $a$, the value of $x_t$ (i.e. $R_R$), the initial gas
temperature, $T_0$, $\alpha~^\prime$, $x_R (= x_t/R_c)$, $P$, the
measured mass loss time scale, $\tau_{loss}$, the ratio $P \Gamma
(x_R, \alpha^\prime)/ \tau_{loss}$, and the equivalent $f$ as derived
from Equation~(\ref{eq:fest}).  In the derivation of $\Gamma$, we set
$\beta=2$, as in Figure~1.  The galaxy parameters are chosen to be
representative of small dwarf galaxies \citep{Mateo98}, and the
variations of the parameters examine the effects of mass, orbital
radius, and gas temperature upon the mass loss rate.  The high
temperatures here and below are chosen to be representative of either
warm neutral, or warm ionized gas with metal cooling.  Throughout the
simulation, the gas temperature remains nearly constant.

\begin{figure}
\plotone{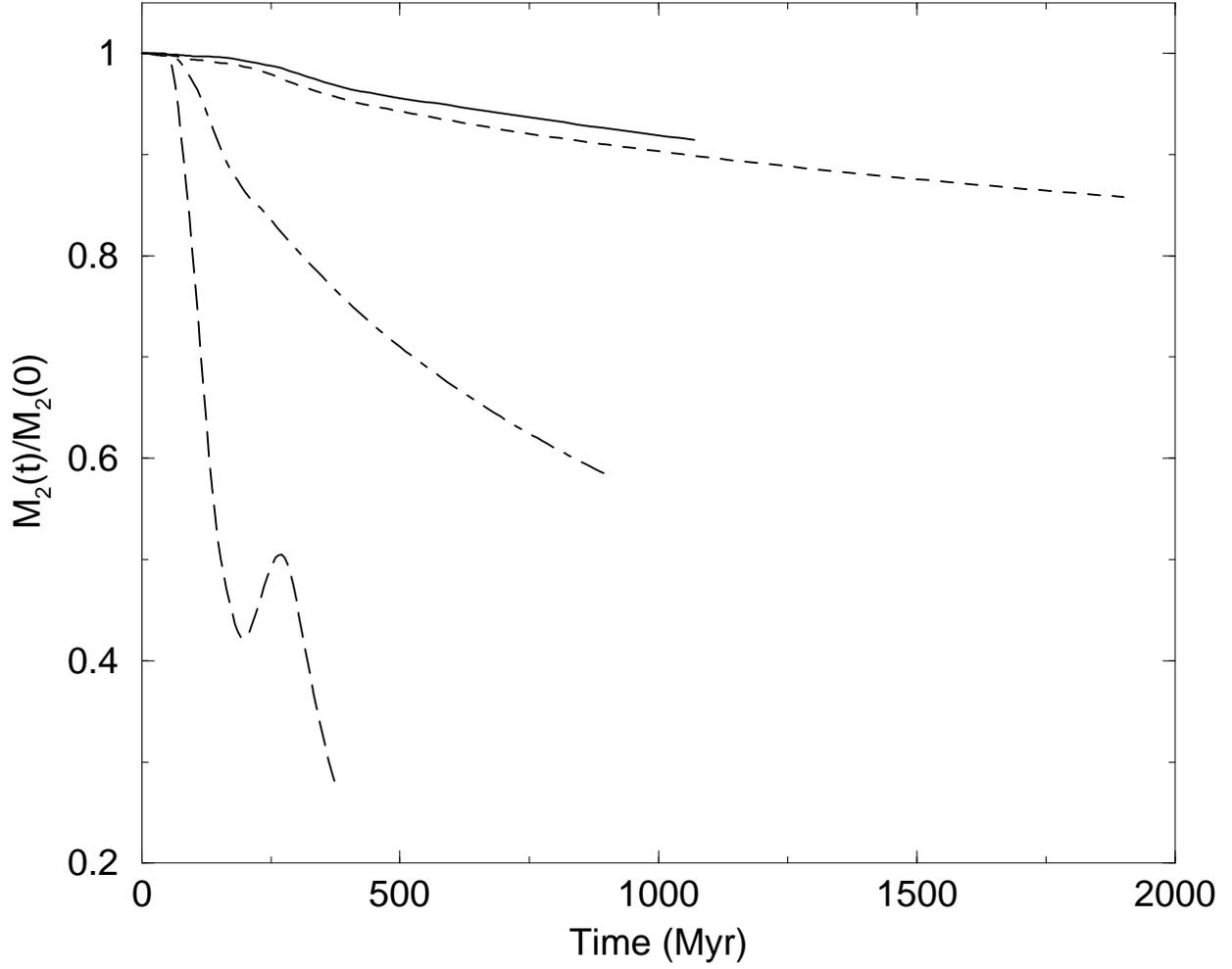}
\caption{The values of $M_2(t)/M_2(0)$ for the three-dimensional
dwarf galaxy models.  The
results for Model~1 are indicated by the solid curve, Model~2 by the
long dashed curve, Model~3 by the short dashed curve, and Model~4 by
the dot-dashed curve.  The
values of $\tau_{loss}$ are found from exponential fits to the curves.
}
\label{fig:3Dgalaxymass}
\end{figure}

\begin{figure}
\epsscale{1.0}
\plotone{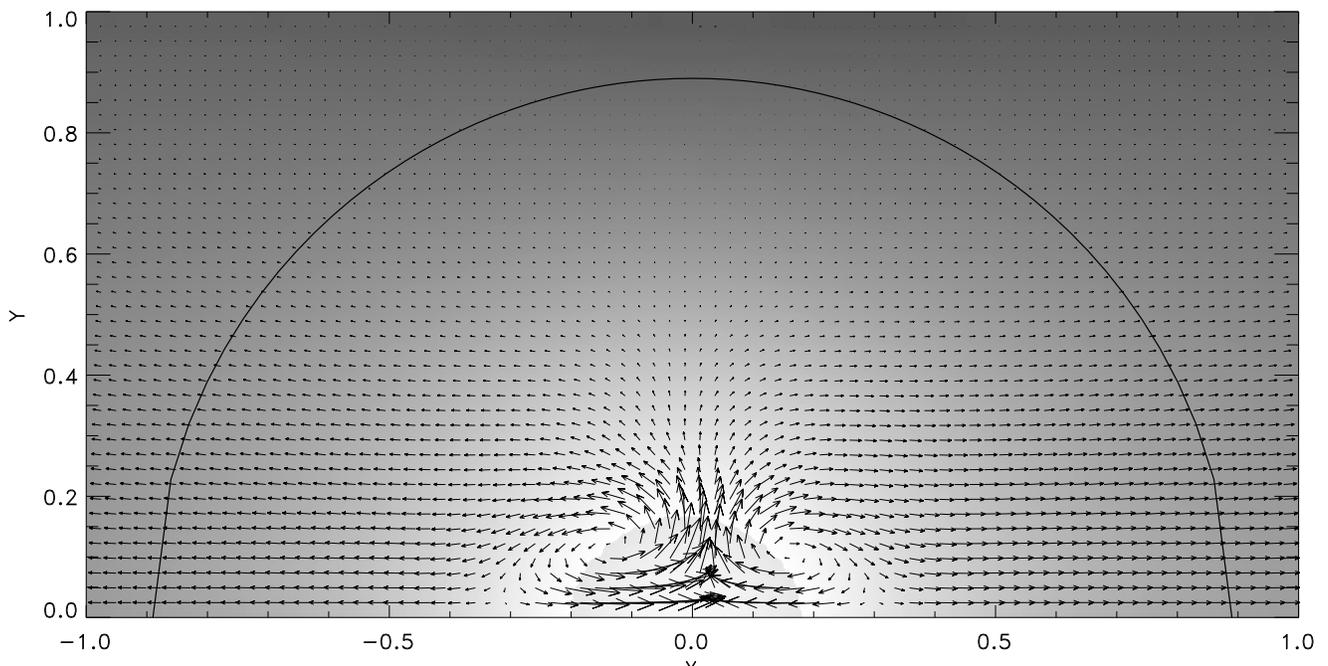}
\caption{The mass flow, $\rho{\bf v}$, for the three-dimensional dwarf
galaxy model 3D-4d at 680~Myr.
}
\label{fig:3Dflow}
\end{figure}

The values of $M_2(t)/M_2(0)$ are shown in Figure~\ref{fig:3Dgalaxymass}.  In
deriving the values of $\tau_{loss}$ quoted in Table~\ref{tab:3Dgalaxies},
the curves have been fitted to the form 
\begin{equation}
M(t)=M(0)\exp\left(-t/\tau_{loss}\right).
\label{eqn:massfit}
\end{equation}
As can be seen from Figure~\ref{fig:3Dgalaxymass} and below, the
curves are not always well represented by exponentials.  It is
apparent, however, that the derived values of $\tau_{loss}$ represent
well the rates at which gas is lost from the systems.  The time scales
for mass loss are comparable to that seen in the one-dimensional case
above.  In comparison with the analytic results in \S2, we find from
Equation~(\ref{eq:f}) that $f \sim 0.2-1.2$ which agrees poorly with that
inferred from Equation~(\ref{eqn:timescale}).  As indicated above, the
departure from hydrostatic equilibrium may be the cause of this
disagreement.  In Figure~\ref{fig:3Dflow}, we plot the mass flow field
($\rho{\bf v}$) for Model~3D-4d at 680~Myr.  The figure shows a slice in $x$
and $y$, at $z=0$.  The greatest values
of $\vert\rho{\bf v}\vert$ occur in the complex flow at the center of the
galaxy.  Away from the center, the vectors of $\rho{\bf v}$ are almost
parallel to the $x$-axis.  The semi-circle in Figure~\ref{fig:3Dflow} has
radius $x_t$, and indicates the size of the Roche lobe in the spherical
approximation (the actual potential used in the model is highly aspherical,
as indicated in Figure~\ref{fig:rochepotential}).  Clearly,
only a fraction of the Roche surface is open to outflow.  The complex
flow within the core is time-dependent, but the flow outside of the core is 
not observed to vary significantly over time.  The largest velocities
present, near the tidal radius, are 10~km~s$^{-1}$.

Also apparent from the differences of $\tau_{loss}$ among the models
is the extreme sensitivity to $\alpha$ (cf.
Equations~\ref{eqn:timescale} and \ref{eqn:gamma}).  The critical value
of $\alpha~^\prime \sim 6$, half the value derived in the analytic
approximation.  For $\alpha<\alpha_1$, a significant fraction of the
original gas in the dwarf galaxy is lost within an orbital period.
The ionized gas in both the lower-mass galaxy and the higher-mass
system at lower orbital radius (Models~2 and 4) is largely lost within
1~Gyr, which would preclude extended epochs of star formation.

\begin{figure}
\epsscale{1.0}
\plotone{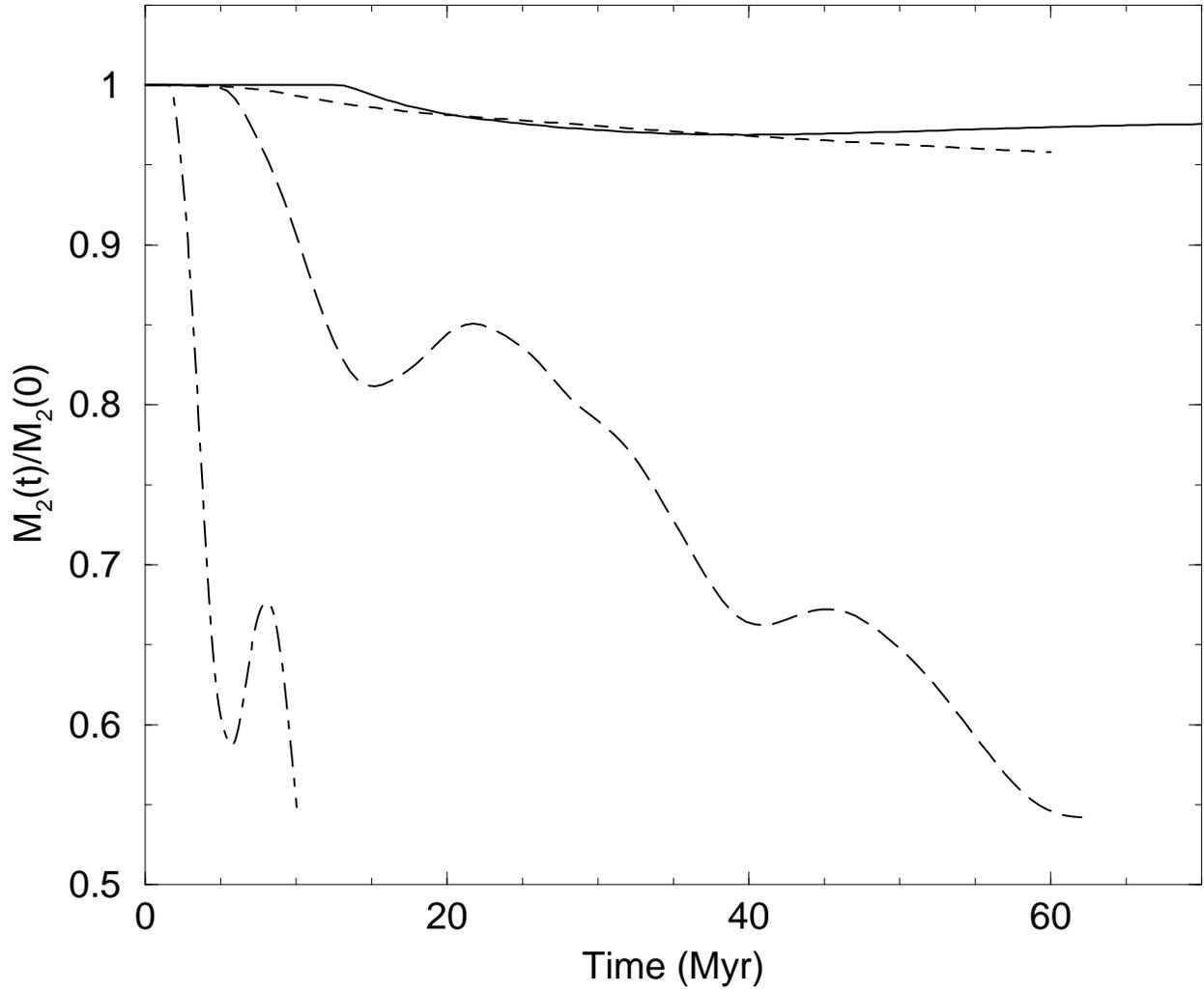}
\caption{The values of $M_2(t)/M_2(0)$ for the cluster models, displayed
as in Figure~\ref{fig:3Dgalaxymass}.
}
\label{fig:3Dclustermass}
\end{figure}

The globular cluster models are listed in Table~\ref{tab:3Dclusters},
which lists the same quantities as in Table~\ref{tab:3Dgalaxies}.  The
masses and core radii are representative of globular clusters
\citep{PM93}.  The orbital radii are either just outside of the Solar
circle, or well within it (most globular cluster orbit closer to the
center of the Galaxy than the Sun).  The evolution of $M_2(t)/M_2(0)$
for the models is shown in Figure~\ref{fig:3Dclustermass}.

As was the case for the dwarf galaxies, ionized gas is lost quickly
from globular clusters, on time scales $\lesssim$~100~Myr.  The
critical value $\alpha^\prime \sim 6$ as for the dwarf galaxy.  Again,
the rapid mass loss induced by tides precludes extended epochs of star
formation in these systems.  Note that for $\alpha^\prime \sim 1$, the
theoretically determined $f$ is much larger than that estimated from
(\ref{eqn:timescale}).  In this limit, the assumption of quasi-equilibrium
breaks down near the Lagrangian point.  Nevertheless, the
dependence of $\tau_{loss}$ on the magnitude of $\alpha^\prime$ is
well established.

\section{Discussion and Consequences for the Evolution of Dwarf Systems}
\label{sec:discussion}

The essence of our results is contained in
Tables~\ref{tab:1Dgalaxies}, \ref{tab:3Dgalaxies}, and
\ref{tab:3Dclusters}.  The results of the calculations in \S\ref{sec:1Dmodels}
indicate that gas in typical dwarf spheroidal galaxies (dSphs) can 
be fully ionized, even by 
a modest background UV flux.  But for the more extended
and dwarf elliptical galaxies, a fraction of the residual gas
can remain neutral at their central regions, especially during more recent
epochs.  Even for these galaxies, mass loss occurs if the dwarf galaxies 
are captured as satellites of larger galaxies or clusters of galaxies.

As can be seen in the tables, and expected from
\S~\ref{sec:analytic}, the time scale for mass loss is extremely
sensitive to the ratio $\alpha\equiv\phi_0/c_s^2$.  The magnitude of
$\phi_0$ can be directly measured from the central velocity
dispersion.  If $\alpha\gg \alpha_1 \simeq 6-10$, then the
gas is tightly held within the potential of the dwarf stellar
system.
The density contrast between the center of the galaxy and the tidal radius,
given by Equation~(\ref{eq:f}),
$\rho(x_t,0,0)/\rho(0,0,0)\sim \exp(-\alpha) \ll 1$.  In this limit,
the time scale for mass loss, $\tau_{loss}\gg\tau_d$, where $\tau_d$
is the dynamical time scale of the dwarf system.  The reverse is true
when $\alpha = \phi_0/ c_s^2 < \alpha_1$.

For many dwarf galaxies, the velocity dispersion is comparable to the
sound speed of ionized gas, and so, if the gas is ionized, the system
should have $\alpha\approx 1$.  Once $\alpha<\alpha_1$, our results
indicate that the gas may be lost to the dwarf system on a time scale
no longer than several times $\tau_d$, which is shorter than both the
orbital period, $P$, and
the time scale on which subsequent generations of stars would be expected to
form \citep{ML93, LM94}.  Multiple possible sources of ionizing
radiation exist, including extragalactic UV, UV from a nearby massive
galaxy, or UV from stars formed within the dwarf galaxy \citep{WHK97,
KBS97, DML03}.  The 
evolution of dwarf galaxies therefore depends upon their proximity to a
massive galaxy, the source of the ionizing radiation, and when or if
the gas within the dwarf galaxies is ionized.

Previous investigation by Kepner et al. (1997) suggest that
the external UV radiation in early epochs can fully ionize residual 
gas in typical-mass dwarf galaxies, and so rapid mass loss might be
expected.  The results in \S{\ref{sec:1Dmodels}} show, however, that hot
ionized gas can be retained in potentials with velocity 
dispersions comparable to or less than the sound speed of the gas, provided
that the host galaxy is isolated.  Thus, even in the presence of an intense
UV background, low-mass dwarf galaxies do not automatically lose their gas
content, albeit the density 
distribution can become relatively flat for systems which contain hot gas.

Such an environmental dependence may help to explain differences between
some types of dwarf galaxies.  Dwarf irregular (dIrr) and dwarf elliptical 
(dE) galaxies are observed to be similar in
terms of their dark-matter potentials and ordinary matter content.  But, a
significant amount of ordinary matter in dIrrs is
in the form of gas, whereas little gas is found in dEs.
It has been suggested that this difference is due to the efficiency of
gas removal (Lin \& Faber 1983), motivated by the fact that dEs
are primarily located in clusters of galaxies,
whereas dIrrs are mostly found in the field.  
Although the presence of hot gas in clusters of galaxies can 
lead to efficient stripping \citep{MWBL93},
the present results indicate that tidal effects can also remove the ionized
gas in dwarf galaxies within a galactic cluster environment
within several Gyr.

At the center of large clusters, however, where the density is nearly
homogeneous or declines slowly with radius, the tidal potential is
compressive rather than disruptive.  In these regions, dwarf galaxies may
also be able to retain their gas content, even if it is mostly ionized.
Nucleated dwarfs are often found in these central regions of clusters of
galaxies.  The retention of residual gas may enable the dwarf systems
to undergo repeated star formation, possibly leading to the formation of
their nucleated structure.  This dependence upon location is in the same
sense as that proposed by Babul \& Rees (1992).  In that earlier work, it
was proposed that the added confinement by the denser gas near cluster
centers would help to prevent supernova-driven mass loss from dwarf
galaxies, allowing them to retain their gas for subsequent generations of
star formation.  Away from the cluster center, however, where tidal forces
are disruptive, they shall act in opposition to external confinement, to
drive mass loss.  In those regions, the presence of a significant external
medium may even enhance mass loss, via the Kelvin-Helmholtz instability
\citep{MWBL93}.

The evolution of non-isolated dwarf systems depends upon both their masses
and their proximity to a massive galaxy.
The results in \S~\ref{sec:3Dmodels} indicate that if a massive galaxy forms
near a dSph prior to the efficient conversion of the internal gas of the dSph
into stars, then the remaining gas shall be rapidly lost to tidal stripping if
it becomes ionized.  In the most extreme case, when background UV heating has
ionized the gas within the dSph, preventing it from forming any stars, tidal
stripping of the gas shall lead to the
formation of a starless, dark-matter mini-halo orbiting the larger galaxy.
Only the gas component of the smaller systems is efficiently lost, because
the outflow is continually replenished by the attempt to re-establish
quasi-hydrostatic equilibrium.
The dark-matter mini-haloes may be preserved, because they were formed in
relatively high density environments \citep{Mooreetal99}.  Thus, the results
presented here naturally imply that the halo of large galaxies may be
populated with many `stripped' dark matter, self-gravitating systems.

By contrast, the gas within the most massive dSphs, and that within dEs 
may be self-shielded, and remain mostly neutral.  That, coupled with their
deeper potentials, allows them to retain their gas for a much longer time
than low mass dSphs, and form multiple generations of stars before losing
their gas, even when in close proximity to a massive galaxy.  The dependence
of the rate of tidal stripping upon the mass of the dwarf system
may provide an explanation for the ``missing satellite galaxy''
problem, because, in accordance with the CDM model, the galactic mass function
decreases with $M_d$.

Tidal loss is also crucial for globular clusters.  As in dSphs,
$\phi_0^{1/2}$ for globular clusters is a few to a few tens of km~s$^{-1}$,
comparable to the sound speed of ionized gas.  
The highly homogeneous chemical abundance among stars within
any given globular cluster \citep{SK77,Cohen79,SK82,RF84,Bolte87a,
Bolte87b} requires thorough mixing of the proto-cluster clouds.  Even with
efficient turbulent mixing, many internal crossing time scales are
required \citep{ML90}.  Thus, prior to the formation of presently observable
stars in globular clusters, gas must be retained for a time comparable
to or longer than their Galactic orbital period.  The baryonic density of
globular clusters is higher than that of dSphs, and so, prior to efficient
star formation, the gas was likely to be self-shielded from external heating.
Because the value of $\alpha_1$ is much larger for cool atomic and molecular
gas, the gas can then be retained for several Galactic
orbits.  Self-gravity of the gas was also probably very important for
globular clusters at early epochs, which would further help them to retain
their gas.

The internal UV production following efficient star formation within a
globular cluster would, however, ionize the gas, after which would be rapidly
lost to the system.  The same fate would apply to the most massive dSphs
orbiting close to a massive galaxy.
In the presence of a tidal field, therefore, star formation within dwarf
systems can become a self-terminating process, even in the absence of ram
pressure stripping, supernova heating, or other means of violent gas
ejection.

In the hierarchical galaxy formation scenario, large galaxies are formed
through coalescence of dwarf galaxies.  Galactic tides 
enable the rapid decoupling of the gas from dwarf
galaxies.  The tidally removed gas is detached gently from the dwarf systems,
and shall form thin tidal streamers, similar to the Magellanic
Stream.  These gas streams are dispersed by differential gravity,
precession, nonlinear shock dissipation, and drag by the residual gas
in the halo.  The tidally removed gas does, however, retain its angular
momentum.  When it is mixed with the residual halo gas, together they
can form the disk \citep{SGV99}.  The ``tidally stripped''
dark matter clumps shall themselves undergo subsequent orbital decay due
to dynamical friction.  In outer regions of the of the large parent galaxies,
the residual gas in the halo is tenuous, possibly allowing the streamers 
to preserve their integrity.  Repeated Galactic encounters may lead to
recapture of the gas from the streamers by the dwarf system, and induce
secondary star
formation.  The follow-up investigation on the tidally removed gas
will be analyzed and presented in subsequent work.

\begin{acknowledgements}
This work was performed under the auspices of the U.S. Department of
Energy by University of California, Lawrence Livermore National
Laboratory under Contract W-7405-Eng-48.  This work is partially
supported by NASA through an astrophysical theory grant NAG5-12151.
\end{acknowledgements}

\clearpage
\begin{deluxetable}{ccccccccccccc}
\tablewidth{0pt}
\tablecaption{One-Dimensional Galaxy Models \label{tab:1Dgalaxies}}
\tablehead{
\colhead{Model} &
\colhead{$M_{d}$} &
\colhead{$r_0$} &
\colhead{$\sigma$} &
\colhead{$\alpha~^\prime$} &
\colhead{$J_\nu$\tablenotemark{a}} &
\colhead{${M_{gi}}\over{M_d}$} &
\colhead{${M_{Hf}}\over{M_{gi}}$} &
\colhead{${R_R}\over{R_n}$} &
\colhead{$P$} &
\colhead{$\tau_{loss}$} &
\colhead{${\tau_{loss}}\over{P}$} &
\colhead{$\Gamma$} \\
\colhead{} &
\colhead{$10^7 M_\odot$} &
\colhead{kpc} &
\colhead{km~s$^{-1}$} &
\colhead{} &
\colhead{} &
\colhead{} &
\colhead{} &
\colhead{} &
\colhead{Gyr}&
\colhead{Gyr}&
\colhead{} &
\colhead{} 
}
\startdata
1D-1 & 3.66 & 1  & 18 &0.14 &40 & 0.1  & 0.15 & 23.0 & 3.85 & 0.7 & 0.18 & 0.06\\
1D-2 & 3.66 & 1  & 18 &0.74 & 1 & 0.1  & 0.74 & 51.0 & 3.85 & 6   & 1.56 & 0.06\\
1D-3 & 3.66 & 1  & 18 &0.93 &0.1& 0.1  & 0.90 & 8.2 & 3.85 & 17  & 4.42 & 0.06\\
1D-4 & 3.66 & 1  & 18 &0.41 &40 & 0.2  & 0.52 & 39.6 & 3.85 & 2.5 & 0.65 & 0.06 \\
1D-5 & 7.32 &1.35&21.6&0.18 &40 & 0.1  & 0.24 & 183 & 5.44 & 2.8 & 0.51 & 0.11
\enddata
\tablenotetext{a}{The units of $J_{\nu}$ are
$10^{-24}$erg cm$^{-2}$ s$^{-1}$ sr$^{-1}$ Hz$^{-1}$} 
\end{deluxetable}

\clearpage
\begin{deluxetable}{cccccccccccccc}
\tablewidth{0pt}
\tablecaption{Three-Dimensional Galaxy Models \label{tab:3Dgalaxies}}
\tablehead{
\colhead{Model} &
\colhead{$M_{d}$} &
\colhead{$R_c$} &
\colhead{$\phi_0^{\prime 1/2}$} &
\colhead{$a$} &
\colhead{$x_t$} &
\colhead{$T_0$} &
\colhead{$\alpha^\prime$} &
\colhead{$x_R$} &
\colhead{$P$} &
\colhead{$\tau_{loss}$} &
\colhead{$\Gamma$} &
\colhead{${P \Gamma}\over{\tau_{loss}}$} &
\colhead{$f$} \\
\colhead{units} &
\colhead{$10^7 M_\odot$} &
\colhead{pc} &
\colhead{km~s$^{-1}$} &
\colhead{kpc} &
\colhead{kpc} &
\colhead{$10^3$ K} &
\colhead{} &
\colhead{} &
\colhead{(Gyr)}&
\colhead{(Gyr)}&
\colhead{} &
\colhead{} &
\colhead{} 
}
\startdata
3D-1d & 1 & 350& 9.2 & 50 & 1.1& 1& 10  & 3.1 & 1.4 & 10  & 10  & 1.4 & 0.17 \\
3D-2d & 1 & 350& 9.2 & 50 & 1.1& 8& 1.3 & 3.1 & 1.4 & 0.3 & 0.04& 0.19 & 1 \\
3D-3d & 10& 500& 26 & 50 & 2.3& 8& 10  & 4.6 & 1.4 & 9.5 & 2   & 0.3 & 0.29 \\
3D-4d & 10& 500& 23 & 20 & 1.2& 8& 8.2  & 2.5 & 0.6 & 1.5 & 3   & 1.2 & 0.16
\enddata
\end{deluxetable}

\clearpage
\begin{deluxetable}{ccccccccccccc}
\tablewidth{0pt}
\tablecaption{Three-Dimensional Cluster Models \label{tab:3Dclusters}}
\tablehead{
\colhead{Model} &
\colhead{$M_{dm}$} &
\colhead{$R_c$} &
\colhead{$\phi_0^{\prime 1/2}$} &
\colhead{$a$} &
\colhead{$x_t$} &
\colhead{$T_0$} &
\colhead{$\alpha^\prime$} &
\colhead{$x_R$} &
\colhead{$P$} &
\colhead{$\tau_{loss}$} &
\colhead{${P \Gamma}\over{\tau_{loss}}$} &
\colhead{$f$} \\
\colhead{} &
\colhead{($10^5 M_\odot$)} &
\colhead{(pc)} &
\colhead{(km~s$^{-1}$)} &
\colhead{(kpc)} &
\colhead{(pc)} &
\colhead{($10^3$K)} &
\colhead{} &
\colhead{} &
\colhead{(Gyr)}&
\colhead{(Gyr)}&
\colhead{} &
\colhead{} 
}
\startdata
3D-1g & 2 & 10 & 9 & 10 & 99 & 1 & 9.4  & 9.9 & 0.29 & 3.2 & 0.07 & 0.77 \\
3D-2g & 2 &  7 & 11& 10 & 87 & 8 & 1.73 & 14  & 0.29 & 0.1 & 0.06 & 1\\
3D-3g & 2 &  7 & 10&  2 & 34 & 1 & 11   & 4.8 & 0.06 & 1.6 & 0.37 & 0.27 \\
3D-4g & 2 &  7 & 10&  2 & 34 & 8 & 1.5  & 4.8 & 0.06 & 0.016 &0.1 & 1
\enddata
\end{deluxetable}

\clearpage

\clearpage

\clearpage

\clearpage

\clearpage

\clearpage

\clearpage

\clearpage

\end{document}